\documentclass[aps,pra,reprint,graphicx,superscriptaddress]{revtex4-1}
\usepackage{graphicx}
\usepackage{subfigure}
\usepackage{hyperref}
\usepackage{amsmath}
\usepackage{amssymb}
\usepackage{epstopdf}
\usepackage{bbm}

\newcommand{\bra}[1]{   \langle #1 |  }
\newcommand{\ket}[1]{  { | #1 \rangle}  }

\begin{document}

\title{Experimental quantum state engineering with time-separated heraldings \\from a continuous-wave light source: a temporal-mode analysis}

\author{K. Huang}
\affiliation{Laboratoire Kastler Brossel, UPMC-Sorbonne Universit\'{e}s, CNRS, ENS-PSL Research University,
Coll\`{e}ge de France, 4 place Jussieu, 75005 Paris, France}
\affiliation{State Key Laboratory of Precision Spectroscopy, East China Normal University, Shanghai 200062, China}

\author{H. Le Jeannic}
\affiliation{Laboratoire Kastler Brossel, UPMC-Sorbonne Universit\'{e}s, CNRS, ENS-PSL Research University,
Coll\`{e}ge de France, 4 place Jussieu, 75005 Paris, France}

\author{V.B. Verma}
\affiliation{National Institute of Standards and Technology, 325 Broadway, Boulder, CO 80305, USA}

\author{M.D. Shaw}
\affiliation{Jet Propulsion Laboratory, California Institute of Technology, 4800 Oak Grove Dr., Pasadena, California 91109, USA}

\author{F. Marsili}
\affiliation{Jet Propulsion Laboratory, California Institute of Technology, 4800 Oak Grove Dr., Pasadena, California 91109, USA}

\author{S.W. Nam}
\affiliation{National Institute of Standards and Technology, 325 Broadway, Boulder, CO 80305, USA}

\author{E Wu}
\affiliation{State Key Laboratory of Precision Spectroscopy, East China Normal University, Shanghai 200062, China}

\author{H. Zeng}
\affiliation{State Key Laboratory of Precision Spectroscopy, East China Normal University, Shanghai 200062, China}

\author{O. Morin}
\thanks{Present address: Max-Planck-Institut f\"{u}r Quantenoptik, Hans-Kopfermann-Str. 1, D-85748 Garching, Germany.}
\affiliation{Laboratoire Kastler Brossel, UPMC-Sorbonne Universit\'{e}s, CNRS, ENS-PSL Research University,
Coll\`{e}ge de France, 4 place Jussieu, 75005 Paris, France}

\author{J. Laurat}
\email{julien.laurat@upmc.fr}
\affiliation{Laboratoire Kastler Brossel, UPMC-Sorbonne Universit\'{e}s, CNRS, ENS-PSL Research University,
Coll\`{e}ge de France, 4 place Jussieu, 75005 Paris, France}
\date{\today}

\begin{abstract}
Conditional preparation is a well-established technique for quantum state engineering of light. A general trend is to increase the number of heralding detection events in such realization to reach larger photon-number states or their arbitrary superpositions. In contrast to pulsed implementations, where detections only occur within the pulse window, for continuous-wave light the temporal separation of the conditioning detections is an additional degree of freedom and a critical parameter. Based on the theoretical study by A.E.B. Nielsen and K. M{\o}lmer in [Phys. Rev. A 75, 043801 (2007)] and on a continuous-wave two-mode squeezed vacuum from a nondegenerate optical parametric oscillator, we experimentally investigate the generation of two-photon state with tunable delay between the heralding events. The present work illustrates the temporal multimode features in play for conditional state generation based on continuous-wave light sources.

\end{abstract}

 
\maketitle

\section{Introduction}

Quantum state engineering of non-classical light is a key ability for applications in quantum information sciences \cite{Illuminati2006}. An efficient method for preparing free-propagating quantum light states is based on a so-called conditional preparation technique. An appropriate measurement on one mode of a bipartite correlated system will project the other mode into a targeted state \cite{Mandel,Dakna1998}. The state preparation is thus probabilistic but heralded. Over the recent years, such technique with single-photon heralding has been successfully used to generate non-Gaussian states based either on pulsed or continuous-wave parametric down-conversion, such as single-photon Fock state \cite{Lvovsky2001,Zavatta2004,Fasel2004,Polzik2007,Mosley2008,Benson2008,Mitchell2011,Morin2012,Fortsch2012,Morin2013a, Fekete2013,Monteiro2014,Luo2015}, optical Schr\"{o}dinger kittens \cite{Ourjoumtsev2006b,Molmer2006,Nielsen2006, Wakui2007,Jove} and recently hybrid entanglement between particle-like and wave-like qubits \cite{Morin2014,Jeong2014}.

The generation of states involving larger photon-number components requires multiple detection events. Pioneering experiments succeeded in using two or thee-photon detections to generate higher photon-number Fock states or their superpositions \cite{Ourjoumtsev2006,Zavatta2008,Bimbard2010,Yukawa2013,Morin2013b}, including larger Schr\"odinger cat states \cite{Ourjoumtsev2007,Namekata2010,Gerrits2010,Etesse2015,Huang2015}. The recent development of high-efficiency superconducting single-photon detectors enables to reach much larger heralding rate, even in the case of multiple conditionings \cite{Huang2015}, and opens up the promise of a variety of novel protocols \cite{Andersen2015}.

In contrast to the pulsed regime where the acceptance window of the heralding events is defined by the pulse temporal profile itself, in the continuous-wave regime these events can occur at different times. Such time separation of the conditioning detections is an additional degree of freedom and can strongly affect the heralded states by introducing a multimode temporal structure \cite{Nielsen2007A,Nielsen2007C,Sasaki2008,Morin2013b}. For example, large-amplitude coherent-state superpositions have been obtained by time-separated two-photon subtraction operated on a continuous-wave single-mode squeezed vacuum \cite{Takeoka2008,Takahashi2008}. Similarly, considering two-mode squeezed vacuum, A.E.B. Nielsen and K. M{\o}lmer have theoretically investigated how the fidelity of the generated states can be affected by the time-separation and have defined optimal temporal modes for Fock-state generation \cite{Nielsen2007C}.  

In this paper, we report on the experimental investigation of this latter scheme with a time-separated conditional detection from a continuous-wave nondegenerate optical parametric oscillator (OPO) operated below threshold. This experiment relies on our recent demonstration of high-fidelity two-photon superposition states where we have shown that a small delay between two conditioning events does not compromise the two-photon fidelity \cite{Huang2015}. Thanks to newly-developed high-efficiency superconducting nanowire single-photon detectors, an unprecedented preparation rate was achieved. Here, this feature enables us to acquire a sufficient amount of data in a reasonable time to cover temporal separation between the two heralding clicks in a range much longer than the width of the temporal mode defined by the OPO cavity. Therefore, we can postselect the temporal delay between triggers within this range, and then explicitly demonstrate the behavior of the resulting state with this delay. This work, in agreement with the seminal theoretical study by Nielsen and M{\o}lmer, illustrates the temporal modal structure that plays a central role in such continuous-wave generation protocols.

The paper is organized as follows. In Sec. \ref{section2} we first provide a basic model of time-separated conditional state preparation in the continuous-wave regime. We then detail the corresponding experimental setup in Sec. \ref{section3}. The conditioning mode features are described. The experimental results are presented in Sec. \ref{section4}, with tunable delay between the two heralding events. We also investigate the specific case where the temporal mode is chosen to be fixed, as defined by one of the two heralding events. Section \ref{section5} concludes the paper.

\section{Model of conditional state preparation with time-separated heraldings}
\label{section2}

In this section, we remind the principles of state generation with time-separated conditionings from a two-mode squeezed vacuum state, as detailled in \cite{Nielsen2007C}. Temporal mode functions are introduced to characterize the generic multimode structure of the heralded state in this continuous-wave scenario. 

\subsection {Typical scheme}

\begin{figure}[t!]
\centerline{\includegraphics[width=0.92\columnwidth]{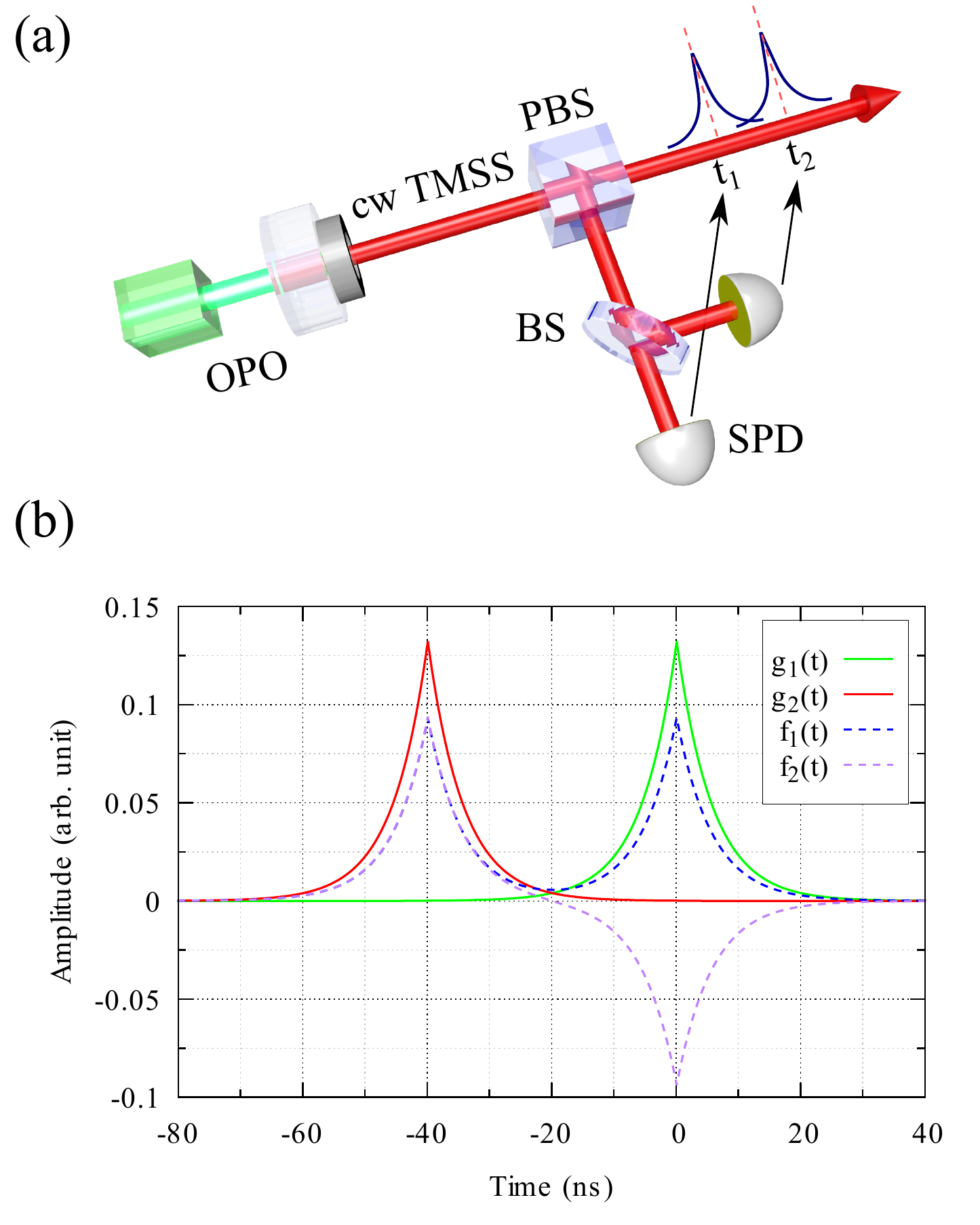}}
\caption{(a) Generation of two-photon state with time-separated conditional detections. OPO: nondegenerate optical
parametric oscillator, SPD: single-photon detector, PBS: polarization beam-splitter, BS: beam-splitter, cw TMSS: continuous-wave two-mode squeezed vacuum state. (b) Temporal mode functions for an OPO cavity bandwidth $\gamma = 53$ MHz. The time separation of the heralding events is set to 40~ns. }
\label{figure1}
\end{figure}

The  generation of two-photon state with time-separated conditional detections is sketched on Fig. \ref{figure1}(a). The initial light source is a continuous-wave two-mode squeezed vacuum state (TMSS) generated for instance by a nondegenerate optical parametric oscillator. The orthogonally-polarized signal and idler modes, which are photon-number correlated, are spatially separated by a polarization beam-splitter. In the ideal case, the simultaneous detection of $n$ photons will project the signal mode into a $n$-photon Fock state. For this study, the conditioning path is split into two parts by a balanced beam-splitter, which are detected by two single-photon detectors. In the following, we consider the general case where the heralding detections have a temporal delay and we investigate how this delay affects the modal structure of the heralded state. 

\subsection {Temporal mode description}

Let us assume that two detection events occur at time $t_{1}$ and $t_{2}$ on the conditioning mode of the continuous-wave TMSS. Due to the finite cavity bandwidth $\gamma$ (FWHM) of the OPO, each photon detection defines a trigger temporal mode function given in the limit of a pump far below threshold by \cite{Nielsen2007C}
\begin{equation}
 \label{Eq:mode1}
{g_i}(t) = \sqrt {\pi \gamma}  {e^{ - \pi \gamma \left| {t - {t_{i}}} \right|}}\ ,
\end{equation}
where $i = 1,2$. 

Accordingly, the heralded two-photon state can be generally written as:
\begin{equation}
\left| \Psi_2  \right\rangle  = \frac{1}{{\sqrt{1 + I^2}}}\iint {d{t}d{t'}{g_1}({t})} {g_2}({t'}){{\hat a}^\dag }({t}){{\hat a}^\dag }({t'})\left| 0 \right\rangle\,
\end{equation}
where ${\hat a}^\dag ({t})$ corresponds to the operator associated with the idler photon in the mode in which the heralding detection took place. $I$ denotes the overlap between the two trigger modes:
\begin{eqnarray}
I &=& \int g_1(t)g_2(t)dt=e^{ - \pi \gamma |t_{1}-t_{2}|}(1 + \pi \gamma |t_{1}-t_{2}|)\nonumber\\
&=& e^{ - \pi \gamma |\Delta t|}(1 + \pi \gamma |\Delta t|)
\end{eqnarray}
with $\Delta t$ the delay between the two events.

Moreover, the state $\left| {{\Psi _2}} \right\rangle$ can be reformulated with two orthogonal temporal mode functions, called symmetric and antisymmetric modes, constructed in the following way:
 \begin{equation}
 \label{Eq:mode2}
 \begin{split}
 {f_1}(t) &= \frac{1}{{\sqrt {2(1 + I)} }}\Big [{g_1}({t}) + {g_2}({t})\Big ]\ , \\ 
 {f_2}(t) &= \frac{1}{{\sqrt {2(1 - I)} }}\Big [{g_1}({t}) - {g_2}({t})\Big ]\ .
 \end{split}
 \end{equation}
Given these two modes, the heralded state can also be expressed as
 \begin{equation}
 \label{Eq:state}
 \begin{split}
 \left| {{\Psi _2}} \right\rangle  &= \frac{{1 + I}}{{2\sqrt {(1 + {I^2})} }} \Big [\int {dt{f_1}(t){{\hat a}^\dag }(t) \Big ]^2} \left| 0 \right\rangle  \\&- \frac{{1 - I}}{{2\sqrt {(1 + {I^2})} }}\Big [\int {dt{f_2}(t){{\hat a}^\dag }(t) \Big ]^2} \left| 0 \right\rangle  \\ 
 &= \frac{{1 + I}}{{\sqrt {2(1 + {I^2})} }}{\left| {2,0} \right\rangle _{1,2}} - \frac{{1 - I}}{{\sqrt {2(1 + {I^2})} }}{\left| {0,2} \right\rangle _{1,2}}\ ,
 \end{split}
 \end{equation} 
where $\left|x,y\right\rangle_{1,2} = \left|x\right\rangle_1 \otimes \left|y\right\rangle_2$ and $\left|x\right\rangle_i$ corresponds to $x$ photons in the mode $f_i(t)$. This expression indicates that the two-photon state with time-separated conditioning is split between the symmetric and antisymmetric modes. 

More precisely, the two-photon fidelity for the mode $f_i(t)$ can be obtained from the norm square of the weight coefficients:
\begin{equation}
\label{Eq:Fidelity2}
F = \frac{{{{(1 \pm I)}^2}}}{{2(1 + {I^2})}}=\frac{1}{2}\pm\frac{I}{1+I^2}
\end{equation}
where $\pm$ corresponds to $f_1(t)$ and $f_2(t)$, respectively. The mode $f_1(t)$ is therefore the optimal mode for maximizing the two-photon state fidelity \cite{Nielsen2007C}. 

The four mode functions $g_1(t)$, $g_2(t)$, $f_1(t)$, and $f_2(t)$ are given in Fig. \ref{figure1}(b) for an OPO bandwidth $\gamma~=~53~$MHz. The delay between two click events is set to 40~ns. These temporal mode functions will be used in our experimental quantum state tomography and the corresponding results will be presented in Sec. \ref{section4}.

\section{Experimental setup}
\label{section3}
\begin{figure}[htb!]
\centerline{\includegraphics[width=0.99 \columnwidth]{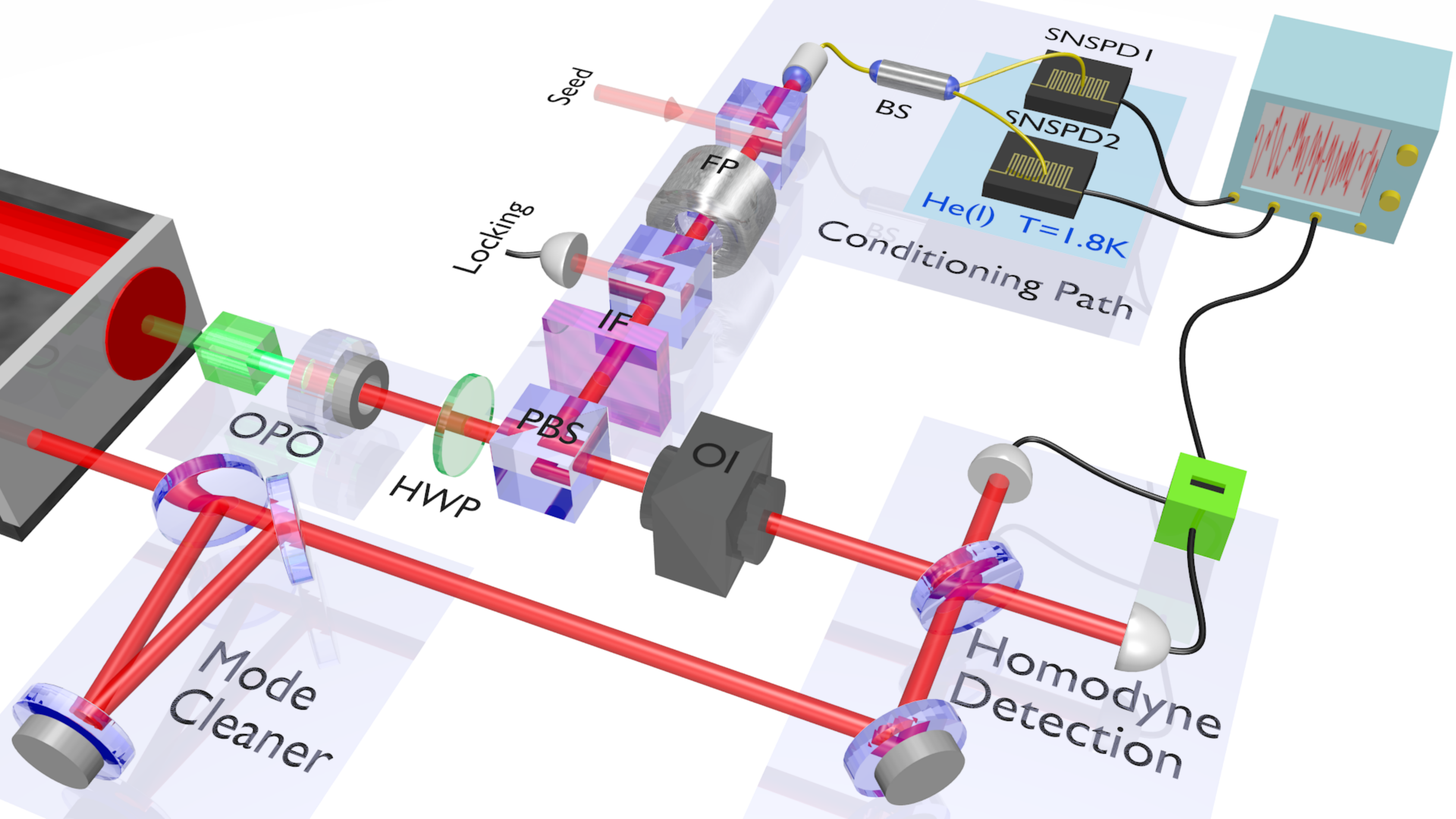}}
\caption{Experimental setup. The initial continuous-wave light source is a two-mode squeezed vacuum state (TMSS) generated by a nondegenerate optical parametric oscillator (OPO) operated far below threshold. The orthogonally-polarized signal and idler beams are separated on a polarization beam-splitter (PBS). The idler is frequency filtered via an interferential filter (IF) and a Fabry-Perot cavity (FP). The conditioning light is  then split on a fiber beam-splitter and the two outputs are detected by high-efficiency WSi superconducting nanowire single-photon detectors (SNSPDs). The time delay between the two triggers is set to be in a user-defined acceptance window (advanced trigger settings provided by a digital oscilloscope). The resulting state is finally characterized by quantum state tomography performed via homodyne detection. The optical isolator (OI) in the signal path is required to prevent the backscattered photons from entering into the conditional path.}
\label{figure2}
\end{figure}

We now turn to the experimental realization. The setup is illustrated on Fig. \ref{figure2}. It consists of three parts: the initial continuous-wave light source, the frequency-filtered heralding path and the heralded state characterization via homodyne detection.

\subsection{Generation of two-mode squeezed vacuum with a type-II OPO below threshold}

A continuous-wave type-II optical parametric oscillator is used to produce photon pairs with orthogonal polarizations. In order to suppress multi-photon components, the pump beam at 532 nm is set far below the OPO threshold (about 2\%).  The OPO is based on a  linear cavity with a semimonolithic configuration. The bandwidth is measured to be $\gamma \simeq$ 53 MHz (FWHM) and the escape efficiency is estimated to be $\eta_\text{OPO} \simeq$ 0.9 \cite{Huang2015,Morin2014}.  More details about the implementation of the OPO have been reported elsewhere \cite{Morin2012,Jove}. 

\subsection{Filtering and characterization of the heralding path}

At the OPO output, the down-converted photons pairs are separated into two spatial modes. The reflected mode is used for the heralding. To remove the non-degenerate modes due to the OPO cavity, the reflected mode is sent through a spectral filtering system consisting of an interferential filter and a Fabry-Perot cavity \cite{Morin2012,Jove,Huang2014}.  The filtered light is then split on a 50/50 fiber beam-splitter and detected by two superconducting nanowire single-photon detectors (SNSPDs) based on tungsten silicide (WSi) \cite{Marsili2013} and optimized at 1064 nm. The system detection efficiency reaches 85\% while the dark count rate is below 10 cps. Note that these unprecedented features are important for our experiment: the high detection-efficiency allows a large preparation rate, which is desirable for experiments involving coincidence detections \cite{Huang2015};  negligible dark noise is also a requisite for achieving high-fidelity state generation \cite{D'Auria2011,D'Auria2012}.

We first investigate the photon-bunching effect in the filtered conditional path. In the limit of low pumping power, the second-order correlation function $g^{(2)}$ is indeed given by:
\begin{equation}
{g^{(2)}}(\Delta t) = 1+{e^{ - 2\pi\gamma \left| {\Delta t} \right|}}{(1 + \pi \gamma \left| {\Delta t} \right|)^2}.
\end{equation}
From this expression, one can notice that the $g^{(2)}(\Delta t)$ parameter increases from 1 to 2 when $\gamma |\Delta t|$ decreases from infinity to zero, which means that the trigger events are bunched in time, as expected for a thermal state. Such bunching effect favors the preparation of two-photon Fock state.

\begin{figure}[t!]
\centerline{\includegraphics[width=0.88 \columnwidth]{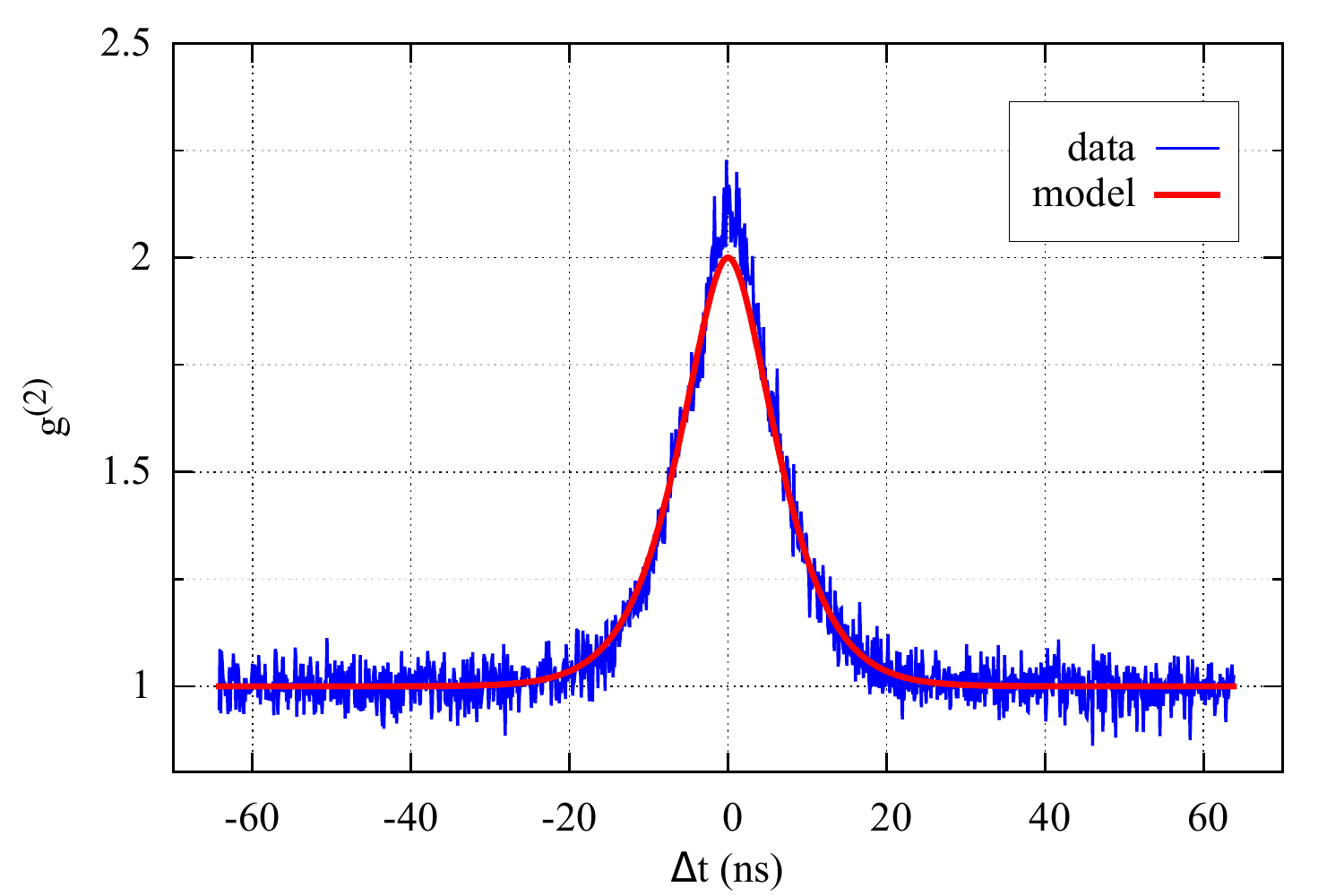}}
\caption{Experimental second-order correlation function $g^{(2)}(\Delta t)$ for the frequency-filtered conditional path. For perfect separation of the signal and idler beams, $g^{(2)}(0)$ is expected to be 2, corresponding to a thermal state.}
\label{figure3}
\end{figure}

In the experiment, we accumulate one million events to obtain a histogram of event counts as a function of the time delay between the two triggers. The histogram is then normalized by the value for large delay as $g^{(2)}(t \to \infty) = 1$. The normalized histogram gives the measured $g^{(2)}(\Delta t)$ in the conditional path, as shown on Fig. \ref{figure3}. The theoretical curve is obtained with the parameter $\gamma$ = 53 MHz. In the ideal case, the state in the conditional path is a thermal state, thus giving $g^{(2)}(0) = 2 $. The slight mismatch of $g^{(2)}(0)$ is due to the imperfect separation of the signal and idler modes. Indeed the two-mode mixing will result in a squeezed state with $g^{(2)}(0)$ larger than 2. Actually, the $g^{(2)}(0)$ value is very sensitive to this mode-mixing and, in the experiment, the two-mode separation is therefore optimized by minimizing this parameter \cite{Huang2015}.

\subsection{Quantum state tomography of the heralded state via homodyne detection}
As the experiment is performed in the continuous-wave regime, the two heralding events can occur at different times. The outputs of the two SNSPDs are connected to a fast digital oscilloscope (Lecroy Wavepro 7300A), offering a dual A-B triggering. The time delay between the coincident triggers is set to be in an acceptance range much longer than the temporal mode duration defined by the OPO cavity bandwidth. 

Given two heralding events, the heralded state is characterized by quantum state tomography via homodyne detection \cite{Lvovsky2009}. The photocurrent $x(t)$ of the homodyne detection is recorded with an oscilloscope at a sampling rate of 10 Gs/s during 500 ns. The local oscillator is swept during the measurements in order to randomize the quadrature phases over the successive acquisitions. As the local oscillator is continuous, post-processing is used to extract the heralded state in a given temporal mode $\xi(t)$. For each realization, we get a single outcome of the quadrature measurement as $x=\int \xi(t) x(t) dt$. In our experiment, one million measurements are accumulated over the 65 ns acceptance window to obtain sufficient quadrature values for quantum state tomography with a maximum likelihood algorithm \cite{Lvovsky2009}.

\section{Experimental quantum state engineering with time-separated heraldings}
\label{section4}

In this section, we present the main results for the conditional preparation of two-photon states with time-separated heralding detections. The two-photon state fidelity depends on the delay and this dependence is characterized here for the limiting cases of very short and very long delay, and for the intermediate general case. Finally, we investigate the specific case of a fixed temporal mode. The degradation of the fidelity for small delays are compared in the two configurations.

\subsection{Limiting case $\Delta t= 0$: generation of two-photon Fock state}
The simplest case corresponds to two conditioning events occurring at the same time, i.e. the ideal situation for the generation of two-photon Fock state. In practice, the heralding triggers are usually accepted in a small coincidence window. In the limit of $\gamma \Delta t \to 0$, one can find from Eq. \eqref{Eq:Fidelity2} that the two-photon fidelity for the optimal temporal mode $f_1(t)$ is given by
\begin{equation}
F \simeq 1 - \left(\frac{\pi\gamma\Delta t}{2}\right)^4\ .
\end{equation}
The optimized fidelity is therefore unity minus a small correction of fourth order in the temporal delay \cite{Nielsen2007C}. A close-to-unity two-photon fidelity can therefore be obtained even though the two detector clicks are not perfectly simultaneous. Practically, one can tune the coincidence window to compromise state preparation rate and fidelity. 

\begin{figure}[t!]
\centerline{\includegraphics[width=0.99\columnwidth]{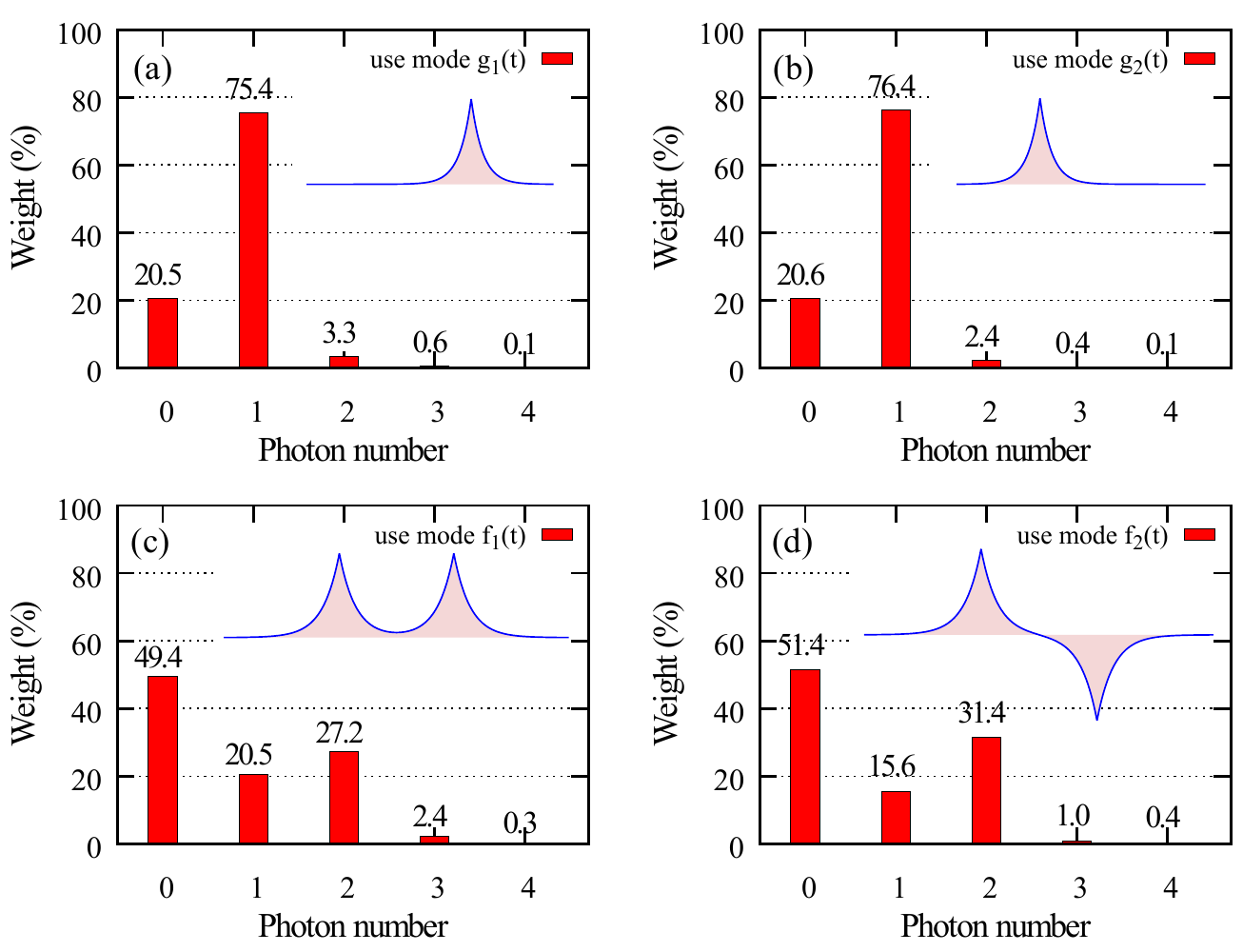}}
\caption{Photon-number distributions of the reconstructed states for the different temporal mode functions $g_1(t)$, $g_2(t)$, $f_1(t)$, and $f_2(t)$ shown in the insets. The distributions are not corrected for detection losses. The delay is set to 40 ns.}
\label{figure4}
\end{figure}

In our experimental realization, we first set the acceptance window to 0.8 ns, much smaller than the typical time given by the inverse of the OPO bandwidth. As a result, the two-photon state fidelity without any loss correction reaches a value as high as 58\%, with a heralding rate of 200 Hz, as reported in \cite{Huang2015}. Corrected for detection losses, this fidelity reaches 79\%, as limited by the square of the OPO escape efficiency. 

\subsection{Limiting case $\Delta t \gg 1/\gamma$: generation of two independent single photons} 

In contrast to the previous case, when the trigger events are very far apart, i.e. $\Delta t \gg 1/\gamma$, the procedure will provide two independent single photons occupying the two temporal mode functions $g_{1,2}(t)$ .

According to Eq. \eqref{Eq:state}, the maximal two-photon state fidelity can be obtained with the symmetric and asymmetric mode functions $f_{1,2}(t)$. As the overlap $I$ goes to zero, this fidelity saturates to 50\%. More generally, when losses are included, the expected two-photon state fidelity in the modes $f_{1,2}(t)$ will be half the value obtained compared to the case $\Delta t = 0$.

To verify these derivations, we experimentally set $\Delta t = 40$~ns, i.e. much larger than the time duration of the temporal mode defined by the OPO cavity. For every pair of clicks with this time separation, the homodyne signal is post-processed with the four aforementioned temporal modes in order to extract the corresponding quadrature amplitudes. Figure \ref{figure4} shows the photon-number distributions of the states reconstructed with the different temporal mode functions $g_1(t)$, $g_2(t)$, $f_1(t)$, and $f_2(t)$, respectively. Due to the losses, the single-photon fidelity is about 76\% for the temporal modes $g_i(t)$. This value gives an expected optimal two-photon fidelity about $0.76^2 \approx 58\%$ when heralded by zero-delay coincident triggers. This value is in agreement with the one measured in the small delay case presented before. This two-photon fidelity also leads to a two-photon fidelity for the case of the $f_{i}(t)$ mode functions of $58\%/2 = 29\%$, which is in good agreement with the measurements given in Fig.\ref{figure4}(c) and \ref{figure4}(d).

\subsection{Intermediate case: transition from single-mode to two-mode temporal structure}

\begin{figure}[t!]
\centerline{\includegraphics[width=0.92 \columnwidth]{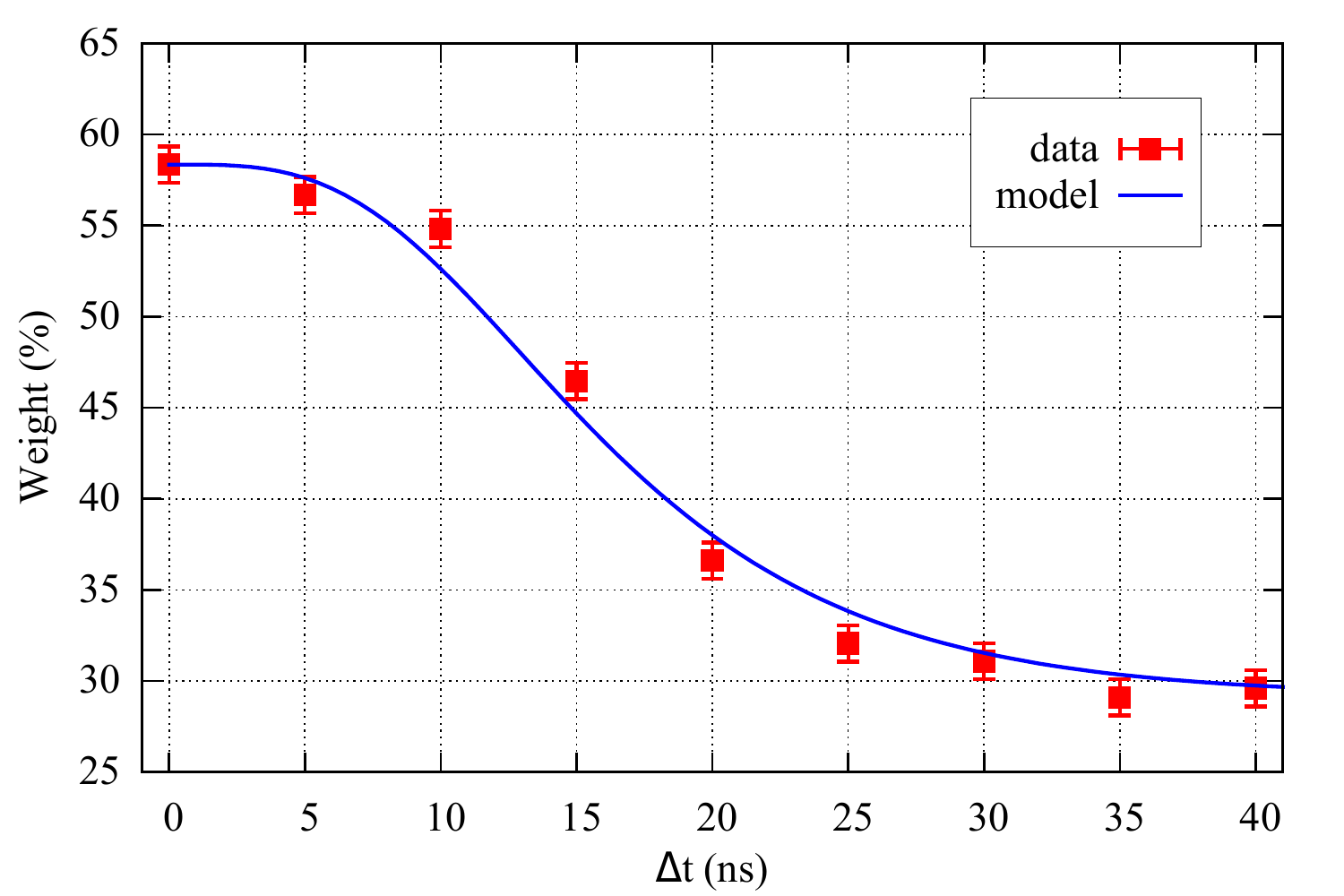}}
\caption{Weight of two-photon component for the optimal temporal mode $f_1(t)$ as a function of the delay between the two trigger. The blue line corresponds to the model taking into account the overall loss and the OPO bandwidth.}
\label{figure5}
\end{figure}

We now investigate the general case with an intermediate delay $\Delta t$. Figure \ref{figure5} shows the two-photon state fidelity obtained when using the optimal temporal mode $f_1(t)$. The degradation of the fidelity with the delay illustrates the transition from single mode to two-mode content. The last data point corresponds to the previous case with large delay.

Interestingly, we can also observe that there is a plateau in the case of small delay, which favors the practical generation of two-photon Fock states. As pointed out before, for small delay, the fidelity only depends on the fourth order in the temporal delay. The fitting line is given by Eq. \eqref{Eq:Fidelity2} by taking into account the overall transmission and detection efficiency  $\eta = 76\%$ and the OPO bandwidth $\gamma = 53$ MHz, which have been measured independently.

The above examples illustrated how the heralded states are affected by using different temporal modes. In particular, we investigated how the two-photon fidelity depends on the delay when using the optimal temporal mode $f_i(t)$ (which depends on $\Delta t$). In the next example, we will consider a fixed temporal mode and show how the reconstructed states are affected by the delay. 

\subsection{Quantum state engineering with a fixed temporal mode}

In real-time quadrature measurement of a two-photon wavepacket, the temporal mode is typically chosen to be fixed \cite{Furusawa2015}. For example, the temporal mode is temporally aligned with one of the detection events \cite{Yukawa2013,Huang2015}.

We first consider a generic mode function $h_1(t)$ as our temporal mode. To derive the photon-number distributions, one can construct a series of orthonormal functions $h_k(t)$ with $h_1(t)$ the first mode. So the heralded state can be rewritten as:
\begin{equation}
\begin{split}
 \left| \Psi_2  \right\rangle  &= \frac{1}{{\sqrt {1+I^2} }}\sum\limits_{m,n} {{C_{mn}}} \Big [\int { {h_m}(t)} {{\hat a}^\dag }({t}) dt \Big ]\\&\Big [\int {{h_n}({t})} {{\hat a}^\dag }({t})dt \Big ]\left| 0 \right\rangle  \\ 
 & = \frac{1}{{\sqrt{1+I^2} }}\sum\limits_{m,n} {{C_{mn}}} \hat A_m^\dag \hat A_n^\dag \left| 0 \right\rangle  \\ 
 & = \frac{1}{{\sqrt {1+I^2} }}\Big [\sum\limits_m {\sqrt 2 {C_{mm}}} {\left| 2 \right\rangle _{{h_m}}} \\&+ \sum\limits_{m > n} {({C_{mn}} + {C_{nm}})} {\left| 1 \right\rangle _{{h_m}}}{\left| 1 \right\rangle _{h_n}}\Big ] \ ,
\end{split}
\end{equation}
where the decomposition coefficients are
\begin{equation}
\begin{split}
 {C_{mn}}   & = \int {h_m^*({t})} {g_1}({t})d{t}\int {h_n^*(t')} {g_2}({t'})d{t'} \ ,\\
 &= {\alpha _m}{\beta _n} \ . 
\end{split}
\end{equation}

If we consider a temporal mode $h_1(t)$ equal to $g_{1}(t)$, which is a practical case, we obtain ${C_{mn}} = {\delta _{1,m}}{\beta _n}$. In this case, the temporal mode $h_1(t)$ must contain at least one single photon. After tracing over other modes, the measurement probabilities for different photon-number states are given by
\begin{equation}
\label{eq:modestwophoton}
{P_2} = \frac{2}{1 + {I^2}}{\left| {{C_{11}}} \right|^2} = \frac{{2{{\left| {{\beta _1}} \right|}^2}}}{{1 + {I^2}}} = \frac{{2{I^2}}}{{1 + {I^2}}}\ ,
\end{equation}
\begin{equation}
\label{eq:modesonephoton}
\begin{split}
 {P_1} & = \frac{1}{1 + {I^2}}{\sum\limits_{m > 1} {\left| {{C_{m1}} + {C_{1m}}} \right|} ^2} \\
 & = \frac{1}{1 + {I^2}}{\sum\limits_{m > 1} {\left| {{C_{1m}}} \right|} ^2} = \frac{1}{1 + {I^2}}{\sum\limits_{m > 1} {\left| {{\beta _n}} \right|} ^2} \\ 
  &= \frac{1}{1 + {I^2}}(1 - {\left| {{\beta _1}} \right|^2}) = \frac{{1 - {I^2}}}{{1 + {I^2}}}\ .
\end{split}
\end{equation}
Since the target mode is always occupied by one heralded single photon, the probability to find no photon in this mode is thus zero, as confirmed by:
\begin{equation}
{P_0} = 1 - {P_1} - {P_2} = 0\ .
\end{equation}

In our experiment, if we use the temporal mode $g_1(t)$ centered at the first detection event, the resulting state is given by
\begin{equation}
\hat{\rho} = P_1 \ket{1}\bra{1} + P_2 \ket{2}\bra{2}\ .
\end{equation}
After taking into account the overall losses on the state (modeled with a fictitious beam-splitter with a power transmittance $\eta$), the state $\hat{\rho'} $ can be written as
\begin{equation}
\begin{split}
\hat{\rho'} &= P_2 \eta ^2 \ket{2}\bra{2} + \Big [2 P_2  \eta (1-\eta)+P_1 \eta \Big ] \ket{1}\bra{1} \\&+\Big [P_2 (1-\eta)^2+P_1(1-\eta)\Big ] \ket{0}\bra{0}\ .
\end{split}
\end{equation}

\begin{figure}[t!]
\centerline{\includegraphics[width=0.97\columnwidth]{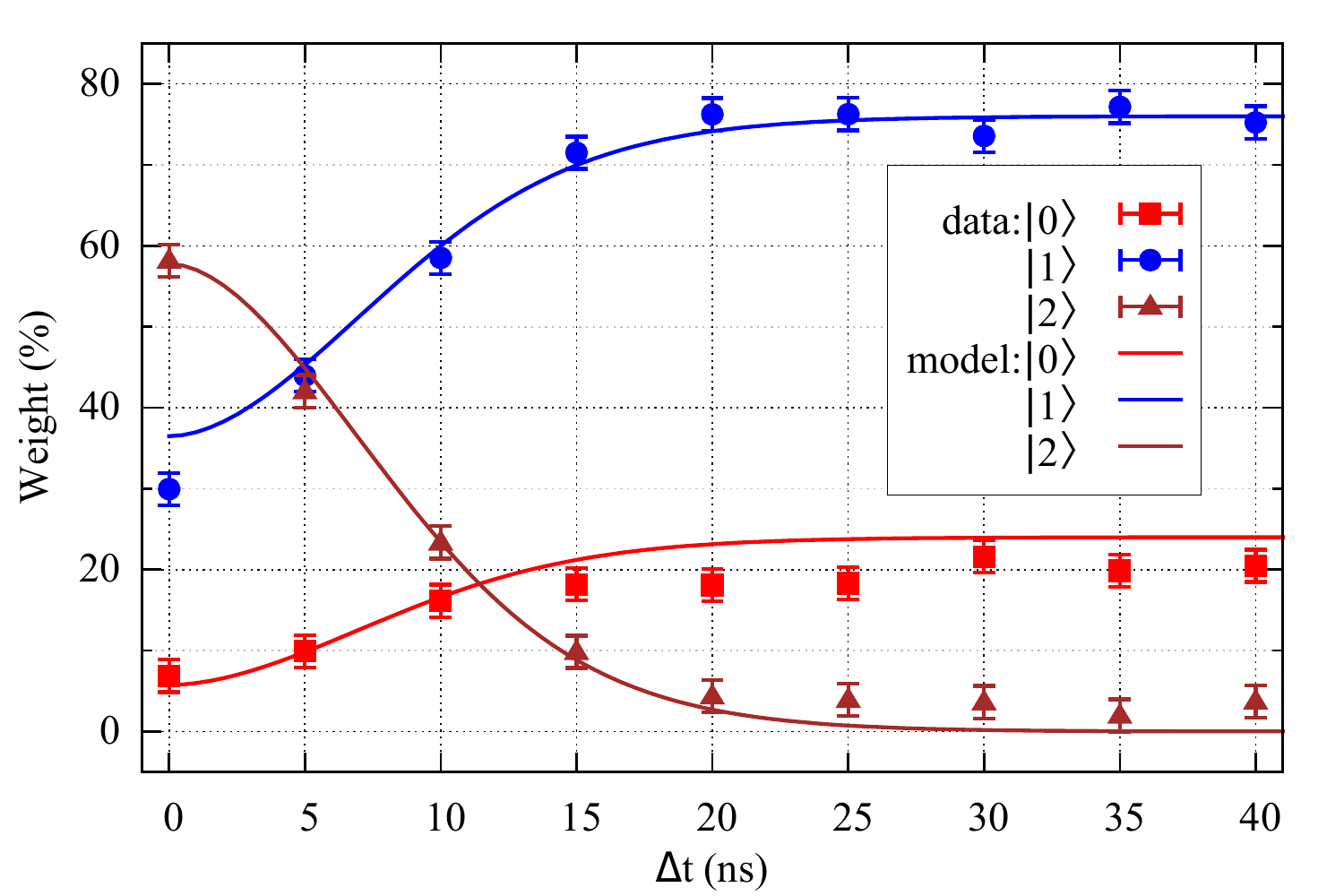}}
\caption{Photon-number weights of the reconstructed states for the fixed temporal mode $g_1(t)$  as a function of the delay $\Delta t$. The two parameters for the model are the OPO bandwidth $\gamma$~=~53~MHz and the overall intensity transmission $\eta  = 0.76\ $.}
\label{figure6}
\end{figure}

Figure \ref{figure6} shows the photon-number distributions for the experimentally reconstructed states with the temporal mode $g_1(t)$ as a function of the delay $\Delta t$ between the triggers. The solid curves are obtained by using the parameters $\gamma$ = 53 MHz and $\eta  = 0.76\ $. The slight discrepancy is due to larger photon-number components, which are not taking into account here and were minimized in the experiment by using a very low pump power. 

As one can notice on Fig.  \ref{figure6}, the decay of the two-photon state fidelity is faster in this case than in the previous study where the optimal temporal mode $f_1(t)$, adapted to each delay, was used. Indeed, using Eq. \eqref{eq:modestwophoton} for small delay, the two-photon state fidelity can be written in the present case as: 
\begin{equation}
F \simeq 1 - \left(\frac{\pi\gamma\Delta t}{\sqrt{2}}\right)^2.
\end{equation}
The fidelity is now unity minus a small correction of second order in the delay.  For a fixed temporal mode, the acceptance window should be thus reduced relative to the optimal adapted case. 

\section{Conclusion}
\label{section5}
In conclusion, we have experimentally investigated the conditional preparation of two-photon Fock states with time-separated heraldings. Due to the continuous-wave light source used here, multiple conditionings introduce a multimode temporal structure. The two-photon state fidelity achieved with the optimal temporal mode has been measured as a function of the delay between the heralding events. Additionally, we studied the practical case where the temporal mode is fixed in the experiment by the first event. For small delay relative to the inverse of the OPO bandwidth, the adapted case leads to a correction of fourth order in the delay while this correction is of second order in the fixed case. This result confirms that the small delay can indeed be used without compromising the state fidelity but should be reduced in the second case. In our two studied cases, the fidelity dropped typically by 5\% for a 10 ns and a 3 ns delay respectively. The present work highlights the importance of temporal modes when working with continuous-wave sources. The subsequent use of the generated state in quantum circuit requires indeed the precise knowledge of the modal structure. We note that efficient methods have been developed recently to experimentally access the optimal mode via raw homodyne data without initial assumptions on the state \cite{Morin2013b,Lvovsky2015}.

\section*{Acknowledgments}
This work was supported by the European Research Council (Starting Grant HybridNet). Part of this research was carried out at the Jet Propulsion Laboratory, California Institute of Technology, under a contract with the National Aeronautics and Space Administration. V.B.V. and S.W.N. acknowledge partial funding for detector development from the DARPA Information in a Photon (InPho) and QUINESS programs. K.H. was supported by the China Scholarship Council. J.L. is a member of the Institut Universitaire de France.

\end{document}